\documentclass{ceurart}

\usepackage{amsmath}
\usepackage{graphicx}
\usepackage{xspace}

\newcommand{\quotes}[1]{`#1'}
\newcommand{\ft}{Feel Transformer\xspace}

\begin{document}

\copyrightyear{2025}
\copyrightclause{Copyright for this paper by its authors.
  Use permitted under Creative Commons License Attribution 4.0}
\conference{}

\title{Transformer-Based Decomposition of Electrodermal Activity for Real-World Mental Health Applications}

\author[1]{Charalampos Tsirmpas}[%
orcid=0009-0000-9110-7716,email=haris@feeltherapeutics.com]
\cormark[1]
\address[1]{Feel Therapeutics Inc., San Francisco, California, US}

\author[1,2]{Stasinos Konstantopoulos}[%
orcid=0000-0002-2586-1726,email=skonstantopoulos@feeltherapeutics.com]
\address[2]{Siglyx, Athens, Greece}

\author[1]{Dimitris Andrikopoulos}[%
orcid=0000-0001-7141-7661]

\author[3]{Konstantina Kyriakouli}[%
orcid=0000-0002-6081-4423]
\address[3]{Johannes Kepler University, Linz, Austria}

\author[1]{Panagiotis Fatouros}[%
orcid=0000-0003-1685-0600,email=panagiotis@feeltherapeutics.com]

\cortext[1]{Corresponding author.}

\begin{abstract}
Decomposing Electrodermal Activity (EDA) into phasic (short-term, stimulus-linked responses) and tonic (longer-term baseline) components is essential for extracting meaningful emotional and physiological biomarkers. This study presents a comparative analysis of knowledge-driven, statistical, and deep learning–based methods for EDA signal decomposition, with a focus on in-the-wild data collected from wearable devices. In particular, the authors introduce the Feel Transformer, a novel Transformer-based model adapted from the Autoformer architecture, designed to separate phasic and tonic components without explicit supervision. The model leverages pooling and trend-removal mechanisms to enforce physiologically meaningful decompositions. Comparative experiments against methods such as Ledalab, cvxEDA, and conventional detrending show that the Feel Transformer achieves a balance between feature fidelity (SCR frequency, amplitude, and tonic slope) and robustness to noisy, real-world data. The model demonstrates potential for real-time biosignal analysis and future applications in stress prediction, digital mental health interventions, and physiological forecasting.
\end{abstract}


\maketitle

\section{Introduction}
\label{sec:intro}

Real-time physiological monitoring using wearable sensors is increasingly recognized as a valuable tool in psychiatric practice, offering continuous, objective data to support mental health assessment and intervention \citep{extra-I, extra-II, extra-III} Electrodermal Activity (EDA), in particular, is a measure of changes in the electrical properties of the skin, influenced by the activity of sweat glands, which are regulated by the autonomic nervous system (ANS). Specifically, EDA reflects the sympathetic branch of the ANS, associated with physiological arousal and responses to emotional or stressful stimuli \citep{boucsein:2012}. When an individual experiences heightened emotional arousal, the sympathetic nervous system stimulates eccrine sweat glands, particularly on the palms and soles, leading to measurable fluctuations in skin conductance \citep{dawson-etal:2007, shields-etal:1987}. These fluctuations are captured as changes in electrical conductance on the skin’s surface, typically measured in microsiemens (uS).

EDA is widely used in psychophysiological research due to its sensitivity to emotional, cognitive, and stress-related processes \citep{rahma-etal:2022,buchwald-etal:2019}. It has been applied extensively in fields such as psychology, neuroscience, and human-computer interaction for applications ranging from stress and anxiety detection to emotion recognition and mental health monitoring \citep{greco-valenza-scilingo:2016}. More recently, measuring EDA has become increasingly accessible with the advent of wearable sensors, allowing continuous and unobtrusive monitoring of physiological arousal in everyday environments. Research-grade devices such as the Feel Monitor \citep{extra-IV}, Empatica E4 \citep{extra-V}, and Shimmer3 GSR+ \citep{extra-VI} have been widely adopted in clinical and research settings \citep{extra-VII,extra-VIII,extra-IX, extra-X,extra-XI, extra-XII}, while consumer-grade devices like the Fitbit and Garmin wearables are increasingly being used in mental health monitoring \citep{extra-XIII, extra-XIV}. These platforms facilitate the detection of stress patterns and physiological correlates of emotional states, facilitating early identification of psychiatric symptoms and supporting timely, personalized interventions \citep{poh-etal:2010}.

EDA signals are typically divided into two primary components: phasic and tonic, each representing different aspects of ANS arousal. The phasic component reflects short-term fluctuations that are closely tied to discrete external or internal stimuli, such as sudden loud noises, visual stimuli, or cognitive tasks that elicit an arousal response. These rapid changes, also known as skin conductance responses (SCR), occur in reaction to specific events and are characterized by sharp, transient increases in skin conductance followed by a gradual return to baseline \cite{boucsein:2012}. Phasic activity provides insights into an individual’s immediate, momentary responses to stimuli, making it useful for studying arousal dynamics, event-related responses, and stimulus-triggered physiological changes \cite{boucsein:2012}.

The tonic component, on the other hand, reflects overall arousal levels over longer periods. It is often referred to as the skin conductance level (SCL) and is indicative of an individual’s general physiological state or emotional baseline \cite{boucsein:2012}. The tonic component changes gradually and is influenced by factors such as stress, attention, or sustained emotional states rather than immediate stimuli. SCL provides essential context for interpreting the individual’s broader arousal state \cite{boucsein:2012}.

Given the distinct physiological information each component provides, it becomes evident that decomposing EDA signals into their phasic and tonic components is critical for a comprehensive understanding of ANS functioning. Given the distinct physiological information each component provides, it becomes evident that decomposing EDA signals into their phasic and tonic components is critical for a comprehensive understanding of ANS functioning. This methodological distinction has direct clinical implications: accurately isolating phasic activity can help identify acute stress responses in real time, while monitoring tonic levels may reveal chronic stress or emotional dysregulation patterns. Such insights can inform timely, personalized interventions in patients with mood or anxiety disorders, making EDA decomposition not just a technical requirement but a clinically actionable step in advancing precision mental health care. By isolating these elements, researchers can better distinguish between momentary, stimulus-induced reactions and the broader, baseline arousal states that evolve over time. 

However, a significant challenge in evaluating EDA decomposition techniques is the lack of a universally accepted ground truth for distinguishing phasic and tonic EDA components, which directly impacts the clinical validation and broader adoption of EDA-based monitoring tools in psychiatric care. Without clear ground truth, it becomes difficult to objectively assess the performance of different decomposition algorithms, which in turn complicates their integration into clinical workflows that require reliability and interpretability. This methodological gap limits the utility of EDA signals in informing treatment personalization, and delays the implementation of real-time stress interventions based on these physiological markers. In lab conditions, researchers trigger ANS responses, which allows them to directly localize SCRs in the signal. Naturally, the duration of such studies is limited. This means that such experiments do not offer the opportunity to observe meaningful changes in the slow-changing SCL. Continuous monitoring, on the other hand, allows the study of SCL but lacks explicit information about where SCRs have occurred.

\citet{bach-friston:2013} also discuss this lack of a definitive reference, and how it hinders the objective assessment of the accuracy and reliability of different methods. They assert, however, that phasic/tonic methods are more accurate than directly analysing the EDA signal as was standard practice previously.

In this article, we investigate whether using deep learning—specifically, a Transformer architecture in a non-autoregressive setting—can be used to address the challenge of EDA decomposition without requiring detailed supervision. This approach enables deployment in wearable devices for continuous, real-world monitoring of autonomic arousal. By facilitating real-time tracking, the method has the potential to enhance clinical decision support systems in psychiatry and bridge the gap between algorithmic innovation and clinical adoption in mental health care. We apply our method to data collected in the wild and compare it against several other methods (both knowledge-driven and statistical). This comparison provides insights into the relative strengths and limitations of conventional knowledge-based domain-specific methods, domain-agnostic methods from the non-parametric statistics literature, and the state of the art in deep learning.

\section{Background}
\label{sec:bg}

In this section, we first present literature specifically targeting EDA signal decomposition. These methods are characterized by the fact that they incorporate extensive domain knowledge on the morphology of SCRs and the EDA signal in general. We then proceed to present generic methods for time series detrending, Transformer neural networks, and, in general, data-driven methods that do not rely on domain-specific priors.

\subsection{Knowledge-Driven Methods}

\citet{benedek-kaernbach:2010a,benedek-kaernbach:2010b}
propose the \emph{Ledalab} phasic/tonic separation method that directly reflects the physiology that generates the EDA signal. They build on earlier works that identify the Bateman biexponential function as accurately reflecting the characteristic steep onset and a slow recovery of phasic impulses; In fact, the Bateman function is both physiologically motivated and consistent with the data \citep{alexander-etal:2005}. In theory, one could simply fit the Bateman function to calculate the phasic component and then simply subtract that from the full EDA signal to calculate the tonic component. However, as noted by \citeauthor{benedek-kaernbach:2010b}, this is not straightforward due to (a) the variability of the two parameters of the Bateman function both across subjects and for a given subject; and (b) the fact that new responses may be superimposed on the recovery slope of previous ones.

What they do instead is to use the Bateman function as a way to detect the segment of the EDA signal that is the result of imposing the response impulse on the underlying skin conductance level \emph{assuming that the impulse peak has been detected.}
In other words, Ledalab does \emph{not} use the Bateman function as a direct detector of SCRs, but as a way to \quotes{guess} a
Bateman-shaped area around each local maximum in the signal.
To account for variation in the parameters of the Bateman function, an optimization task is performed for each maximum to estimate the parameters that maximize metrics inspired from known properties of the phasic and tonic components, but also includes elements that have been empirically found to work.
This is followed by \emph{deconvolution} of the EDA signal over the Bateman function. This gives a \emph{driver} signal: A signal such that its convolution with the Bateman function would give the original EDA signal; or, in other words, a signal where occurrences of the shape of the Bateman function stand out more clearly. Naturally, because of the optimization step, practically all maxima fit the Bateman shape and appear in the driver signal. To separate actual impulses from noise artefacts, Ledalab applies an empirical threshold to identify \quotes{significant peaks}.
Once significant peaks are detected, the region of the EDA signal
that corresponds to each peak (as per the Bateman function) is
zeroed out. The core idea is that phasic impulses are contained in
time, so if we remove the regions with phasic activity, what
remains is a purely tonic signal. The tonic component of the
phasic-activity regions can then be accurately interpolated between the purely tonic regions. Once the complete tonic component is estimated, then simple subtraction gives the phasic component.

\citet{greco-etal:2016} follow a more direct approach in
their \emph{cvxEDA} algorithm. Instead of Ledalab's multi-step method, cvxEDA directly optimizes the complete formulation of EDA as the sum of a Bateman-shaped phasic component, a relatively smooth (cubic spline) tonic component, and some residual (sampling or modelling error). The optimizer finds the best parameters for all components, again guided by priors about the shape of phasic impulses.

\citet{hernando-gallego-etal:2018} follow a similar approach,
also jointly optimizing the parameters of the phasic and the
tonic model. When compared to cvxEDA, one difference is that
their \emph{sparsEDA} algorithm uses a \emph{dictionary}
of impulse models to select from, instead of allowing the
optimizer to set the parameters of the Bateman function.
Also, the method is more explicitly formulated to take into consideration the sampling rate and other theoretical and practical considerations regarding the sampled and discretized signal.

In a more recent variation of this optimization-based line of research, \citet{faghih-etal:2019} use the \citet{zdunek-cichocki:2008} method for finding sparse overlapping signals. However, this method needs to be parameterized with the onset and recovery times, which are known to vary between subjects and even between experiments with the same subject. In order to estimate these parameters, \citeauthor{faghih-etal:2019} first apply cvxEDA
on a short portion of the data.

\citet{jain-etal:2017} emphasise treating a more complex tonic component than the cubic spline interpolation that is used in the works presented above. \citeauthor{jain-etal:2017} note that in the case of wearable devices, the baseline can have abrupt shifts due to movement, of a magnitude that cannot be captured by the usual noise models. In order to represent such discontinuities, they model EDA as being composed of a step baseline function, a tonic component, and the phasic impulses. Exploiting reasonable assumptions regarding the maximum frequency of discontinuities and the almost-sparse nature of the phasic component, they are able to formulate a solvable optimization task.

\subsection{Data-Driven Methods}

The methods presented above are specifically designed for decomposing EDA signal, and thus encode prior knowledge about the shape of SCR. However, the manifestation of SCRs in actual data is not uniform and clear to the extent that we can consider recognizing them a solved problem. The amplitude of the actual responses varies, and the relative amplitude with respect to noise even more so; Furthermore, overlapping SCRs might obscure the slow-recovery pattern predicted by the Bateman function.

Given the above, it makes sense to also experiment with data-driven methods to identify responses as events that stand out from the overall signal without referring to prior observations about their shape. Framing EDA decomposition in this way, it makes sense to also consider the wide variety of statistical \emph{detrending} methods that have been proposed to separate fast-moving events from any slow-moving trend in the background. Detrending typically works by fitting a linear (or, in any case, relatively simple) model for the overall trend and then subtracting that from the signal. Fitting a linear model to a time series is one of the most common tasks in non-parametric statistics, and many textbook methods (such as least squares and the Theil–Sen estimator) are part of daily practice in econometrics, physics, meteorology, and practically every natural science.

Similarly to the statistical methods discussed above, machine learning methods also have the advantage that they can exploit data to find regularities for which we lack a clear definition.
\emph{Artificial Neural Networks (ANN)}
\citep{goodfellow-etal:2016} are known to be universal function approximators with minimal requirements for prior knowledge, but specific network architectures are best suited for different kinds of tasks. In our case, we are looking for a network that is able to capture the positional dependencies that model the steep onset and slow recovery of phasic impulses while simultaneously abstracting away the specific position of each impulse in the EDA signal. The \emph{convolutional neural network (CNN)} architecture
\citep[Ch.~9]{goodfellow-etal:2016} has been specifically designed
to capture such local patterns and, even closer to our case,
\emph{recurrent neural networks (RNN)}
\citep[Ch.~10]{goodfellow-etal:2016} and
\emph{long short-term memory (LSTM)} \citep{hochreiter:1997}
target capturing local events in time series.

Unlike non-parametric statistics, however, machine learning relies on a training dataset where relatively short (although not necessarily exact) spans of signal are annotated as positive and negative examples of the pattern of interest. Manually creating such a dataset at the scale required for training an ANN is a daunting task, and such methods are typically used in tasks where datasets of positive and negative examples are readily available or can be extracted at scale. In our case, it is preferable to focus on \emph{unsupervised} methods that can be trained to separate the fast-moving and the slow-moving component of the EDA signal without explicit examples.

The use of CNNs, designed to reconstruct and/or forecast time series, in an autoregressive and therefore unsupervised manner, can be especially enhanced by using dilated convolutions that are able to capture long-range dependencies effectively along the temporal dimension \citep[closing paragraph]{hewamalage-etal:2021}. Similar successful approaches include WaveNet, a CNN-type model with dilated convolutions for fixed-length time series \citep{borovykh:2017}, and the combination of multiple CNNs, each designed to capture either closeness pattern or long and short-range dependencies in time series \citep{wang:2019}. 

Naturally, in order to be able to distinguish the two components,
signal data points that are far removed from each other must be
taken into account together. A \emph{transformer} model with self-attention provides attention connections with a very wide receptive field where temporal correlations are less locally-focused and more widely connected, allowing it to uncover long-range dependencies \citep{vaswani:2017}. 

A natural issue that arises with using attention on long time series is computational efficiency. In the traditional approach of
scaled dot product attention, the computational complexity and memory requirements scale quadratically over the size of the input, making it difficult to handle inputs beyond 512 tokens \citep{devlin:2019}. At the sampling rate of 8Hz, which is generally accepted as sufficient for capturing EDA features, this restricts our input to 1min. Various modified attention mechanisms have been proposed to tackle this problem, and among the most notable and successfully used are the Informer model, which uses prob-sparse self-attention (provides a probabilistic approach to sparsifying the self-attention matrix) \citep{zhou:2021}, the Reformer (LSH attention by approximating the full self-attention mechanism with a more computationally manageable variant) \citep{kitaev:2020}, and the Autoformer model \citep{wu-xiao-etal:2021}, which uses an auto-correlation mechanism instead of traditional attention.

Our reasons for choosing the Autoformer model are mainly two.
First, the model is very efficient in uncovering long-range dependencies because instead of computing pairwise attention weights between all elements in the sequence, the autocorrelation attention mechanism leverages the statistical properties of time series data to identify and focus on the most predictive parts of the sequence \citep{wu-xiao-etal:2021}. 
Second, the Autoformer is a detrending model, as its blocks sequentially remove and refine a trend component from the time series input. Given the above, we expect the Autoformer to be an appropriate basis for developing a trend-sensitive decomposition mechanism that separates the slow-moving tonic component from the fast-moving phasic component.

\section{Comparing EDA Decomposition Methods}
\label{sec:comparison}

In order to compare algorithms, we would normally go back to their purpose and 
define metrics that reflect how well each algorithm serves this purpose.
This is less straightforward in multi-step methods where we do not have ground
truth annotations for the outputs of the intermediate steps but only the system
as a whole. Nevertheless, the core intuition behind decomposition is that once
the slow-moving SCL component has been removed, SCR peaks will have similar
amplitude so that peak-detection methods can be directly parameterized.

We should here note these characteristics of the SCL and SCR components
are encoded in the optimization step of Ledalab and cvxEDA, for perfectly
noiseless signal, the result is expected to be optimal. However, in data
collected in the wild, either algorithm can fail under different circumstances:
Ledalab optimizes each peak separately, which might result in fitting peaks
that are noise artefacts; cvxEDA optimizes globally, which might result
is counter-intuitive SCL as it tries to fit noise artefacts.
We will revisit this point in Section~5, but what is important is that
the most informed methods in the domain often given different results,
without any clear, automated way to decide which is is more accurate.

Without access to golden-truth decomposition, we need to establish measures
of similarity between decompositions and compare the behaviour of the different
methods. Such a measure of similarity could be the RMS between SCL curves
or that the same SCR peaks are identified under the same peak detection parameters.
However, we observe that the ultimate goal is not the identification of SCR peaks per se, but statistics known to be characteristic of ANS responses: SCR frequency, mean amplitude, and the power spectral density of different frequency bands. In other words, it can happen that two alternative decompositions of the same signal might appear different when comparing whether the same SCRs have been identified, but give the same or very similar feature values.

\citet[Table~2]{kreibig:2010} lists the EDA features proposed in the relevant literature. There seems to be a consensus on the direction of change of SCL, the frequency and amplitude of SCR, and the nonspecific skin conductance response rate. Other features that appear prominently, although not universally, include the Ohmic Perturbation Duration index (OPD) and SYDER skin potential forms. We refer the reader to \citet{silva-etal:2017} for precise definitions. Among those, only the direction of change of SCL, the frequency of SCR, and the amplitude of SCR can be extracted from all decomposition methods, since the other features require recognizing the complete SCR and not only its peak.

Based on the above, in the experiments we present below we use these three
features (direction of change of SCL, frequency of SCR, amplitude of SCR) as
similarity metrics.

\section{Experimental Setup}
\label{sec:exp}

\subsection{Datasets and Methods}

Our dataset is extracted from data acquired in previous studies by
Tsirmpas et al. \citep{tsirbas-etal:2022,tsirbas-etal:2023}, comprising longitudinal EDA data collected over a 16-week period from 40 individuals. This dataset is considerably larger-scale than most datasets previously used to study EDA, both with respect to the number of different users and with
respect to the duration of EDA signal per user. It is also acquired from wearable devices in the wild, offering a very representative sample of what EDA looks like in personalized health and well-being use cases, as opposed to controlled-environment clinical studies and research.

For our empirical comparisons, we have used the following implementations of relevant methods presented in Section~\ref{sec:bg}:
\begin{itemize}
\item \textsl{Ledalab} as implemented by
  Pypsy~0.1.5,\footnote{Available
    from \url{https://github.com/brennon/Pypsy}}
  with the change that the input is resampled to 24Hz instead
  of 25Hz. This has no impact on the method, but makes it easier
  to undersample back to the 8Hz rate of our data.
\item \textsl{cvxEDA} and \textsl{sparsEDA} as implemented in
  NeuroKit2~0.2.10.\footnote{Available
    from \url{https://github.com/neuropsychology/NeuroKit}
    and PyPI. Note that this version or later must be used, as earlier versions do not include our sparsEDA patch.}
\item \textsl{Thiel detrending} as implemented in
  SciPy~1.11.\footnote{Package \texttt{scipy.stats.theilslopes}
    from \url{https://github.com/scipy} and PyPI.}
\end{itemize}

All methods were applied on non-overlapping 3-min frames to produce the tonic component, noting that Theil detrending produces a slope/intercept pair, which is interpreted as a straight line with those parameters. The phasic component was then calculated as the residue from the EDA signal.

For all methods, the EDA signal was first cleaned with a
low-pass filter with a 3Hz cutoff frequency and a 4th order
Butterworth filter. The number of peaks was calculated using
the NeuroKit2 peak detection algorithm with default parameters.
Since we are interested in observing differences between the
peaks detected after different decomposition methods, the
specific algorithm and parameters used are not expected to
affect our results, as long as the same algorithm/parameters
is used for all methods. For this reason, we have \emph{not}
used the peaks reported by decomposition methods such as Ledalab
where detecting SCR is an integral part of the algorithms.

\subsection{Feel Transformer}
\label{sec:feel_transformer}

We base our Feel Transformer on the Autoformer \citep{wu-etal:2021}, but we effected several changes in the architecture, although we largely re-used the implementations of the individual
layers from the original Autoformer implementation. \footnote{Available from
  \url{https://github.com/thuml/Autoformer}}
We used one encoder and two decoder blocks, with the aim of adding
more processing power and depth to our network. At the same time,
using the input sequence again before the decoder blocks acts as a
residual connection. We use the network in a non-autoregressive fashion. That is, there is no left-to-right processing as is usual with the decoder part of a transformer, but rather the sequence is processed bidirectionally in both encoder and decoder blocks.

The rationale behind these changes is that our objective was not prediction (forecasting) as in the original Autoformer paper, but the non-autoregressive reconstruction of the input into two components that, when added, reconstruct the input. The loss function is the mean squared error between the original time
series and the reconstruction.
To bias the decomposition towards a slow-moving SCL and a
fast-moving SCR component, we simplified the network that produces the
SCL part of the output to a max-pooling layer, forcing the deeper part
of the network to reconstruct the SCR component.
The intuition behind this is that the SCR network will be challenged
to converge when it is presented with data where similar morphologies
appear at different amplitude levels; but it can converge by learning
morphologies that are at roughly the same amplitude level, since the
loss is calculated after adding the SCL and the SCR components.

The network of the Feel Transformer has the following characteristics:
\begin{itemize}
\item There is one encoder layer and two decoder layers.
  The dimension of the embedding is 32.
  The output dimension of the first convolution of the
  feed-forward block is 16.
\item There are 4 attention heads.
\item The SCL component is modeled by 1-D average pooling.
\end{itemize}
We tested three different sizes for the average pooling kernel:
(a) 8x60+1 (we will refer to this as \textsl{Feel Transformer~1} in the results
presented below), (b) 8x30+1 (we will refer to this as \textsl{Feel Transformer~2}),
and (c) 8x1+1 (\textsl{Feel Transformer~3}).
Since our data is at 8Hz, this effectively means that the granularity of the SCL
is 60sec, 30sec, or 1sec, respectively.

\section{Experimental Results and Discussion}
\label{sec:results}

\subsection{Analysis of Feature Values}
\label{sec:results:features}

As argued in Section~\ref{sec:comparison}, the most useful features that can be extracted from the decomposed EDA signal are the direction (falling, stable, or rising) of the slope of the tonic component and the density and amplitude of peaks in the phasic component.

We calculated the tonic slope in $\mu S/sec$ by simply dividing
the difference between the last and the first sample in each
frame over its duration (3min). Since the tonic slope can never
be literally zero, we have binned the slopes into three bands, which we called "falling" (most negative slopes), "stable",
and "rising". The bin boundaries are set at 
$-0.001\mu S/sec$ and $0.001\mu S/sec$.
These boundaries were selected because they give a uniform histogram when bundling together the tonic components from all decomposition methods. We then calculate the histograms separately for the tonic components produced by each method.
By comparing these histograms against a uniform
33\% -- 33\% -- 33\% split, we can see any possible bias towards exaggerating or understating the tonic slope.
As we can see in Table~\ref{tab:tonic}, Theil and \ft give results similar to each other and different from the results given by Ledalab, cvxEDA, and sparsEDA, which characterize more frames as \emph{falling}.

Regarding peak density and amplitude, these are given in
Tables~\ref{tab:peak_density} and~\ref{tab:peak_amplitude}.
As can be seen, Ledalab, cxvEDA, detrending, and the first two
configurations of the \ft generally agree on peak density.
When compared on peak amplitude, cxvEDA stands out in giving
considerably higher amplitudes, followed by Ledalab
at around 0.2~$\mu S$, and sparsEDA, detrending, \ft~1 and~2
at around 0.1~$\mu S$.

Ledalab appears to be giving the best quality of results. As far as we can understand, without a specific ground truth label for individual peaks, the Leadalab algorithm makes peaks more
pronounced as evidenced by (a) the higher mean amplitude for roughly the same peak density and (b) the higher fraction of peaks in the 0.2 -- 0.4 $\mu S$ band. On the other hand, the high mean amplitude in the cvxEDA peaks is an artefact of the cvxEDA optimization that we will revisit below.

\begin{table}[b]
\centering
\caption{Histograms of falling/stable/rising slopes for the
  tonic components produced by each decomposition method.
  The three \emph{Feel Transformer} models refer to the
  three configurations detailed in Section~\ref{sec:feel_transformer}.}
\label{tab:tonic}
\begin{tabular}{ccccc}
\hline
Decomp.  & \multicolumn{3}{c}{Tonic Slope}  \\
Method   & Falling & Stable & Rising        \\ 
\hline
Ledalab  & 36\% & 32\% & 32\%            \\
cvxEDA   & 38\% & 30\% & 32\%           \\
sparsEDA & 38\% & 35\% & 27\%           \\
Theil    & 34\% & 34\% & 32\%           \\
\hline
Feel Transf. 1 & 34\% & 32\% & 34\% \\
Feel Transf. 2 & 34\% & 32\% & 34\% \\
Feel Transf. 3 & 34\% & 31\% & 35\% \\
\hline
\end{tabular}
\end{table}

\begin{table}[bp]
\centering

\caption{Mean number of peaks and histograms of the number of peaks
  detected per 3min frame of the phasic component produced by each decomposition method.
  The three \emph{Feel Transformer} models refer to the
  three configurations detailed in Section~\ref{sec:feel_transformer}.}
\label{tab:peak_density}
\begin{tabular}{llrrrrrrrr}
\hline
\multicolumn{2}{l}{Decomp.}
  & Mean &\multicolumn{7}{c}{Histogram of number of peaks}  \\
\multicolumn{2}{l}{Method}
  & Peaks& 0--4 & 5--9 & 10--14 & 15--19 & 20--24 & 25--29 & 30-- \\
\hline
\multicolumn{2}{l}{Ledalab}
 & 11.5 & 16\% & 28\% & 27\% & 15\% & 7\% & 3\% & 2\% \\
\multicolumn{2}{l}{cvxEDA}
  & 11.8 & 18\% & 27\% & 24\% & 16\% & 9\% & 3\% & 3\% \\
\multicolumn{2}{l}{sparsEDA}
 &  9.7 & 23\% & 35\% & 22\% & 11\% & 5\% & 2\% & 1\% \\
\multicolumn{2}{l}{Theil}
 & 10.1 & 21\% & 35\% & 22\% & 12\% & 5\% & 2\% & 2\% \\
\hline
Feel  &1& 10.5 & 19\% & 34\% & 24\% & 13\% & 6\% & 2\% & 2\% \\
Trans-&2&11.2 & 16\% & 31\% & 26\% & 14\% & 7\% & 3\% & 2\% \\
former&3& 26.3 &  7\% & 13\% & 13\% & 12\% &10\% & 8\% &35\% \\
\hline
\end{tabular}

\caption{Mean amplitude and amplitude histogram of the peaks detected
  in the phasic component produced by each decomposition method.
  The three \emph{Feel Transformer} models refer to the
  three configurations detailed in Section~\ref{sec:feel_transformer}.}
\label{tab:peak_amplitude}
\begin{tabular}{llrrrcrcrcrcr}
\hline
\multicolumn{2}{l}{Decomp.}
 & Mean && \multicolumn{9}{c}{Histogram of peak amplitude}  \\
\multicolumn{2}{l}{Method}
 & Ampl.&&  & \hskip -2em 0.005 &  & \hskip -1.5em 0.10 && \hskip -1.5em 0.20 && \hskip -1.5em 0.40  \\
\hline
\multicolumn{2}{l}{Ledalab}
  & 0.196 & & 36\% && 16\% && 12\% && 30\% && 6\%  \\
\multicolumn{2}{l}{cvxEDA}
  & 0.661 & & 45\% && 11\% && 10\% && 26\% && 8\%  \\
\multicolumn{2}{l}{sparsEDA}
  & 0.116 & & 45\% && 13\% && 12\% && 26\% && 4\%  \\
\multicolumn{2}{l}{Theil}
  & 0.109 & & 49\% && 13\% && 12\% && 23\% && 3\%  \\
\hline
Feel  & 1&0.103&& 49\% && 14\% && 12\% && 23\% && 3\%  \\
Trans-& 2&0.102&& 50\% && 14\% && 12\% && 22\% && 3\%  \\
former& 3&0.017&& 86\% &&  5\% &&  3\% &&  6\% && 1\%  \\
\hline
\end{tabular}

\caption{Histogram of the amplitude ranges (max-min) of the
  phasic component produced by each decomposition method.
  The three \emph{Feel Transformer} models refer to the
  three configurations detailed in Section~\ref{sec:feel_transformer}.}
\label{tab:phasic_range}
\begin{tabular}{llrcrcrcrcrcrcrcr}
\hline
\multicolumn{2}{l}{Decomp.}
 & \multicolumn{13}{c}{Histogram of phasic amplitude ranges}  \\
\multicolumn{2}{l}{Method}
 &  & \hskip -2em 0.02 &  & \hskip -1.5em 0.05 && \hskip -1.5em 0.10 && \hskip -1.5em 0.50 && \hskip -1.5em 1.0 && \hskip -1.5em 10 \\
\hline
\multicolumn{2}{l}{Ledalab}
  & 48\% && 16\% && 12\% && 15\% && 3\% && 5\% && 1\%\\
\multicolumn{2}{l}{cvxEDA}
  & 48\% && 17\% && 11\% && 14\% && 3\% && 6\% && 1\%\\
\multicolumn{2}{l}{sparsEDA}
  & 44\% && 18\% && 13\% && 16\% && 3\% && 7\% && 1\%\\
\multicolumn{2}{l}{Theil}
  & 43\% && 18\% && 13\% && 17\% && 3\% && 5\% && 1\%\\
\hline
Feel  & 1 
  & 45\% && 18\% && 12\% && 16\% && 3\% && 5\% && 1\%\\
Trans-& 2& 46\% && 18\% && 12\% && 15\% && 3\%  && 5\% && 0\%\\
former& 3& 82\% &&  8\% &&  4\% &&  5\% && 1\%   && 1\% && 0\%\\
\hline
\end{tabular}

\end{table}

The third \ft configuration uses the smallest kernel and is expected to produce a tonic component that closely follows the EDA curve; the results show that this result is \emph{not} desired as it gives a phasic signal that is of too low amplitude, with many small peaks that are more likely to reflect noise than SCR.

\subsection{Analysis of Phasic Morphology}
\label{sec:results:morphology}

We will now proceed to extract more statistics and to study characteristic examples in order to better understand the
results presented in Section~\ref{sec:results:features}.

\begin{figure}[t]
\centering
\includegraphics[width=\linewidth]{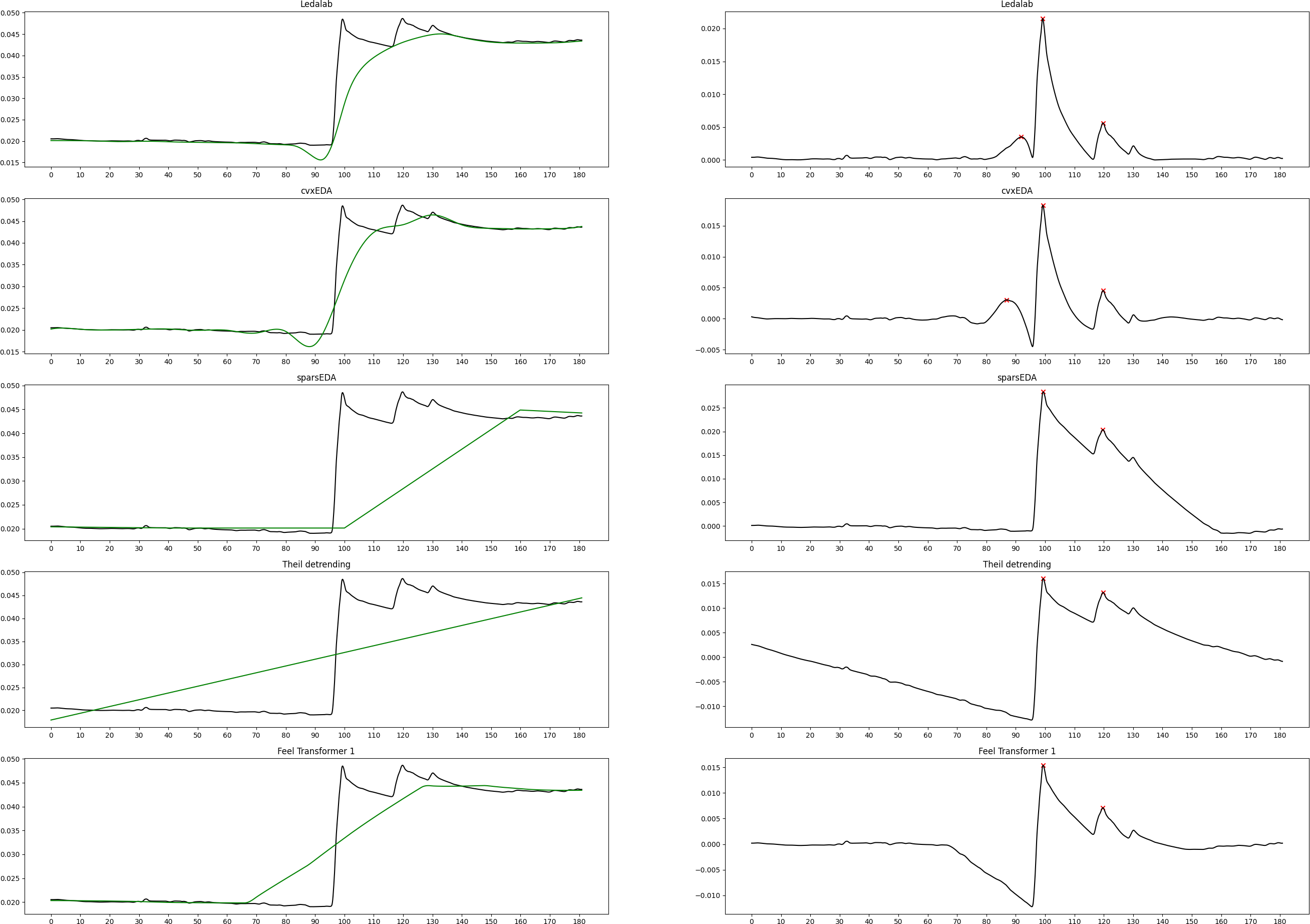}
\caption{A characteristic example of false peaks created by
  the forced smoothness of cubic interpolation even when the
  original signal has abrupt changes.
  EDA signal (black) tonic component (green) are shown on the
  left-hand side; the phasic component and the peaks detected
  are shown on the right-hand side.}
\label{fig:plot1}
\end{figure}

Since the phasic component is the residue between the EDA signal and the tonic component, its range (max-min amplitude) informs us about the complexity of the tonic component: A more complex tonic model will tend to follow the EDA signal more closely and leave a smaller range to the phasic component. The validity of this interpretation of amplitude range is corroborated by the ranges of the three \ft models, where the drastically smaller kernel of the third model indeed gives a significantly narrower value range for the phasic component
(Table~\ref{tab:phasic_range}).

\begin{figure}[t]
\centering
\includegraphics[width=\linewidth]{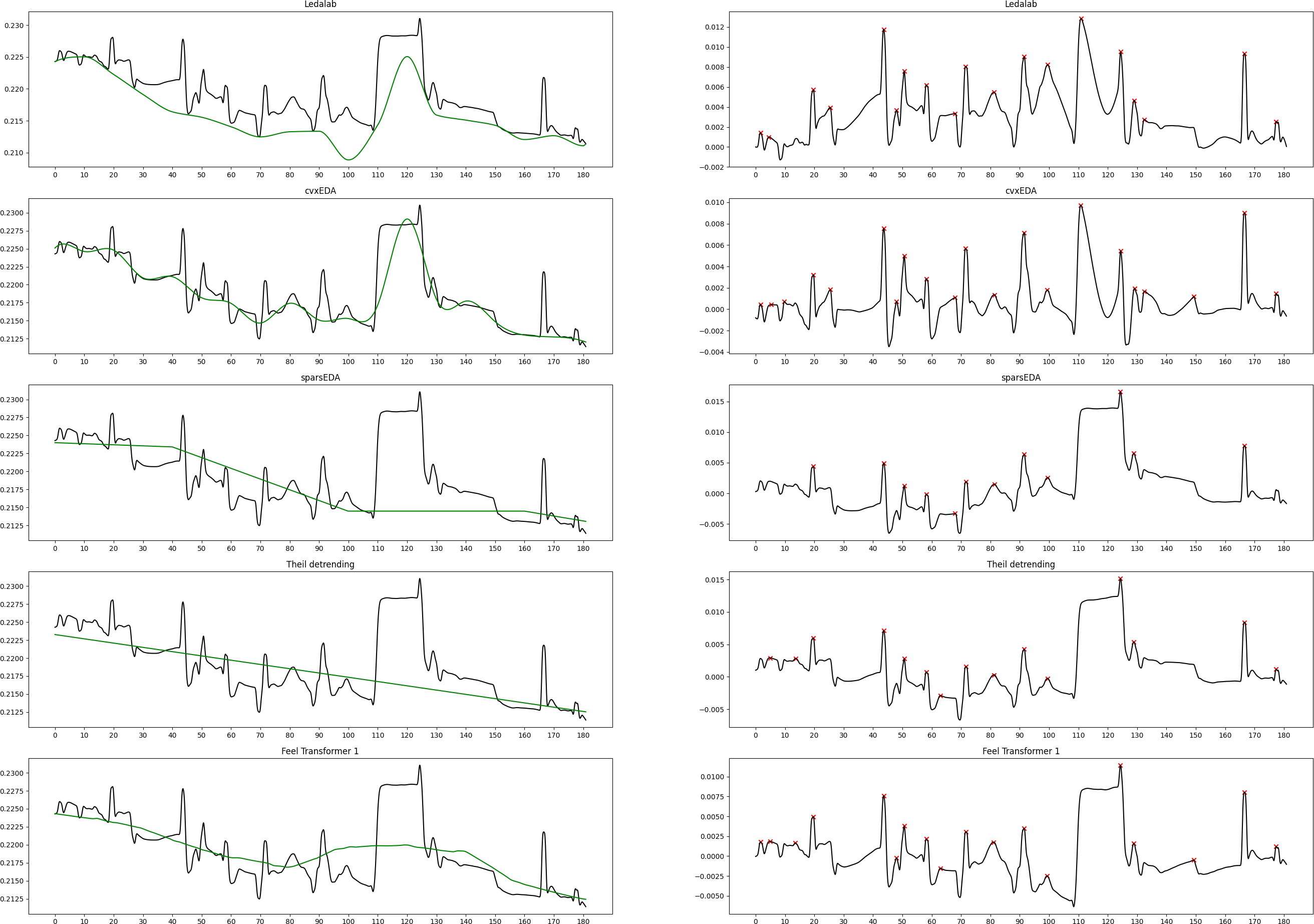}
\caption{A characteristic example of false peaks created by
  cvxEDA.
  EDA signal (black) tonic component (green) are shown on the
  left-hand side; the phasic component and the peaks detected
  are shown on the right-hand side.}
\label{fig:plot2a}
\end{figure}

Coming back to the relatively high mean amplitude of the peaks, it is explained by phenomena such as those observed in Figure~\ref{fig:plot1}. Notice the dip around 90sec in the Ledalab and cvxEDA tonic components, which does not reflect a real drop in the amplitude of the EDA signal but is a side-effect of cubic interpolation. This dip causes the minor amplitude rise at 85sec to become more significant than it really is and to be falsely recognized as a peak. Although in this example both Ledalab and cvxEDA exhibit this phenomenon, it is more common in cvxEDA, possibly because it jointly optimizes the Bateman parameters and the SCL interpolation, affording it more flexibility to \quotes{discover} instances of the Bateman function. Figure~\ref{fig:plot2a} has a characteristic example: using cubic interpolation to treat the sudden change in SCL between 110 and 125sec causes both Ledalab and cvxEDA to create an
artificial peak at 110sec; Note, however, that creates another artificial peak at 150sec by raising the SCL at 140-145sec without having any reason to do so based on the original EDA, so that the falling SCL at 140-150sec creates the rise needed in the phasic component at 145-150sec to recognize a peak.

\begin{figure}[t]
\centering
\includegraphics[width=\linewidth]{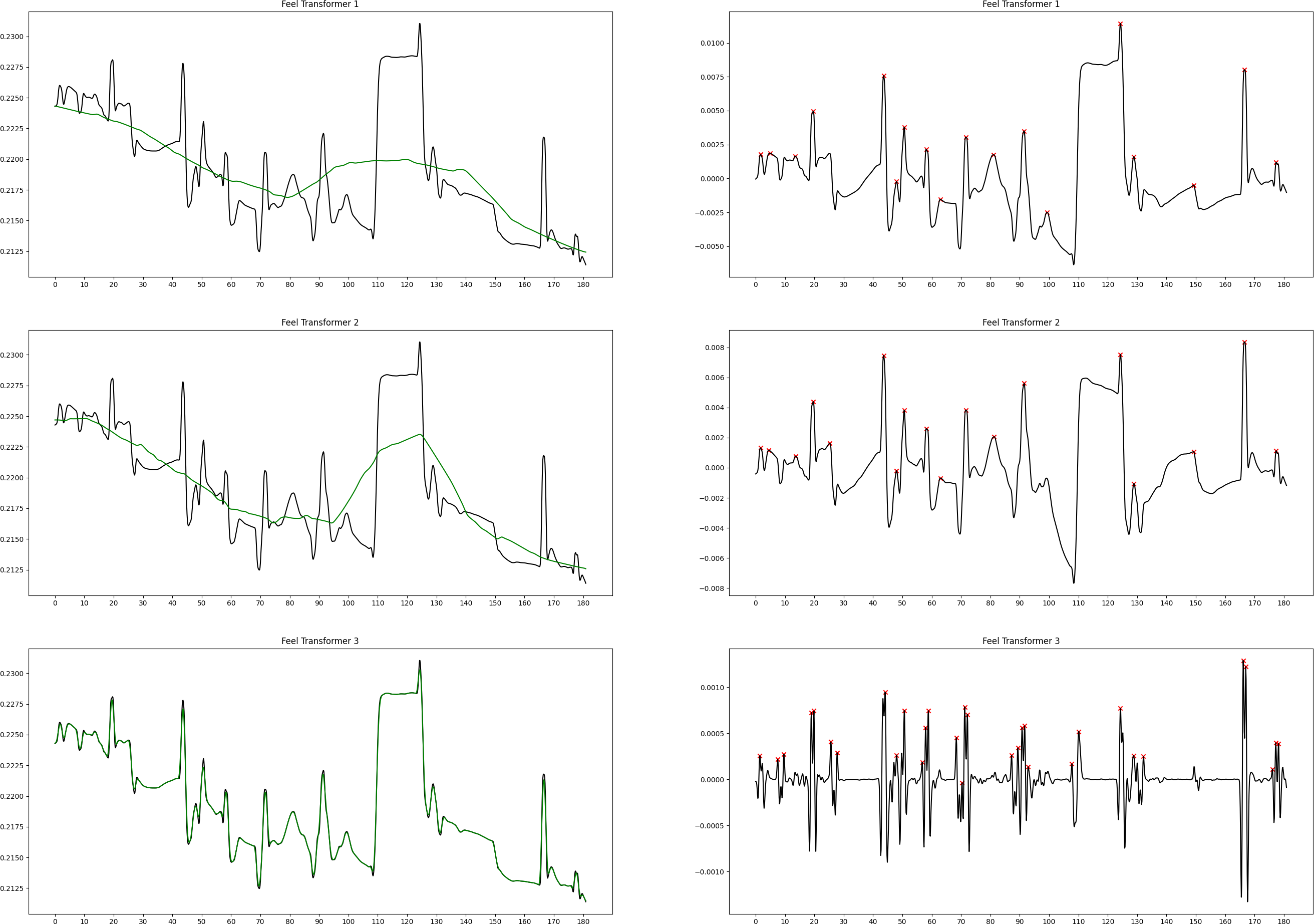}
\caption{Comparison of the three configurations of the
  Feel Transformer.
  EDA signal (black) tonic component (green) are shown on the
  left-hand side; the phasic component and the peaks detected
  are shown on the right-hand side.}
\label{fig:plot2b}
\end{figure}

When comparing Ledalab and cvxEDA with sparsEDA and detrending,
we observe that the latter completely miss the change in SCL level
between 110 and 125sec. Although this is obviously wrong, it does
\emph{not} have any side-effects on the overall tonic slope or 
peak density. It does, however, have a negative impact on peak amplitude, which is higher than it should be. The Feel Transformer
compromises between the two, and falls into the same pitfall,
although for a different reason: The Transformer is not trying to
jointly optimize the SCL and the SCR as cvxEDA does, but it is
trying to minimize the loss measured as the distance from the EDA signal. So it creates an SCL that is falling steeply, trying to reach the lower SCL level after 215sec, and this steep SCL slope creates the rise in phasic at 140-150sec that creates the peak.
When using a finer kernel to allow the SCL to more closely follow the EDA, the problem persists, and when the kernel gets too fine (the third configuration), the results break down completely
Figure~\ref{fig:plot2b}.

\section{Conclusions and Further Work}
\label{sec:outro}

We presented the Feel Transformer, a new method for decomposing EDA signals into a slow-moving tonic component and a sparse, fast-moving phasic component. Our method is based on the Autoformer, a variation of the Transformer NN architecture, but departs from the standard Autoformer in that it explicitly encodes the knowledge that one of the two components must be a relatively simple and slow-moving curve that fits the general trend of the signal. In this respect, the Feel Transformer is more
similar to domain-agnostic detrending, as both estimate the tonic component from the overall signal (including the SCRs), under the (reasonable) assumption that SCRs are sparse and not expected to drastically affect the estimation of the tonic component. On the other hand, the three EDA-specific methods use a three-pass approach where they first use prior knowledge of the overall shape of SCR to identify possible SCRs, then estimate the tonic component from the remaining non-SCR signal only, and then
extract the phasic component and apply peak detection to identify actual SCRs.

The empirical results validate this similarity: When extracting the features generally considered as most useful in EDA analysis (SCL direction, SCR density, SCR amplitude)
\ft agrees with detrending on all three features, while \ft and detrending agree with Ledalab, cvxEDA on SCR density, and with sparsEDA on SCR amplitude (Section~\ref{sec:results:features}). The visual inspection of characteristic frames reveals that the cubic interpolation used by Ledalab and cvxLeda creates non-existent peaks when the EDA signal has abrupt changes. Such changes are rare in laboratory conditions where datasets are usually acquired, but are a lot more common in personal healthcare and well-being applications that rely on the EDA signal acquired in the wild from wearable devices.
It should also be noted that laboratory studies validate methods by observing signal segments that are known to contain ANS responses because such responses have been explicitly elicited by stimuli. However, such approaches cannot validate the tonic
component, since their limited duration cannot contain meaningful SCL changes. Based on the above, we observe that the agreement of \ft with domain-specific methods on the SCR features is a positive indication for the validity of our method; While the minor disagreement between \ft and domain-specific methods on the SCL should \emph{not} be taken into account since the performance of domain-specific methods on the SCL features has not been validated.

As future work, we plan to investigate the robustness of the \ft to noisy signals. Wearable biosensors used in real-world settings often introduce artifacts due to movement, sweat, and intermittent signal loss. In contrast to traditional statistical deconvolution methods---which treat all deviations symmetrically and lack contextual understanding---Transformer-based models can learn to identify and ignore non-informative segments by capturing global signal structure. This ability is particularly critical in differentiating emotional stress from physical exertion in ambulatory monitoring scenarios such as Post-traumatic stress disorder (PTSD) or Attention-deficit/hyperactivity disorder (ADHD). Demonstrating that the Feel Transformer can reliably extract meaningful features from such noisy, in-the-wild data would mark a major advancement over laboratory-optimized methods.

A further advantage of \ft is that, by contrast to the other methods, it is a generative model. This offers the opportunity to 
non-autoregressively simulate or forecast future physiological states (features) and then use the generated signal to extract \emph{further} features besides the ones used to generate the signal in the first place. This enables applications such as anticipatory interventions and anomaly detection in
psychiatry—ranging from predicting stress overload and panic attack onset to forecasting depressive episode relapse.

Recent work supports the feasibility of such applications: Yang et al. \citep{YANG2024100464} demonstrated the use of Transformer-based models to forecast affective states by integrating wearable sensor data with self-reported diaries, achieving high accuracy in mood prediction across temporal windows. Also works like the one from Halkiopoulos and Gkintoni \citep{electronics14061110} highlighted that Transformer and reinforcement learning are used extensively over biosingals like heart rate and EDA fluctuations—during virtual reality–based therapeutic stimuli where predictive simulations enable adaptive adjustments to therapeutic content in real time, tailoring intervention intensity to the individual's physiological profile. The above are some of the studies that highlight the potential of models like the \ft not only to decompose biosignals, but also to generate plausible physiological trajectories—an essential capability for real-world mental health applications.

\end{document}